\documentclass[aps,prl,twocolumn,showpacs,groupedaddress,preprintnumbers,amsmath,amssymb]{revtex4}

\usepackage {amsmath,amssymb,epsfig,color}
\usepackage {graphicx}
\usepackage {dcolumn}
\usepackage {bm}

\begin{document}

\title{Electronic theory for itinerant in-plane magnetic fluctuations in Na$_x$CoO$_2$}

\author {M.M.~Korshunov $^{1,2}$}
 \email {maxim@mpipks-dresden.mpg.de}
\author {I.~Eremin $^{2,3}$}
\author {A.~Shorikov $^4$}
\author {V.I.~Anisimov $^4$}
 \affiliation {$^1$ L.V. Kirensky Institute of Physics, Siberian Branch of Russian Academy of Sciences, 660036 Krasnoyarsk, Russia}
 \affiliation {$^2$ Max-Planck-Institut f\"{u}r Physik komplexer Systeme, D-01187 Dresden, Germany}
 \affiliation {$^3$ Institute f\"{u}r Mathematische und Theoretische Physik, TU Braunschweig, 38106 Braunschweig, Germany}
 \affiliation {$^4$ Institute of Metal Physics, Russian Academy of Sciences-Ural Division, 620041 Yekaterinburg GSP-170, Russia}

\date{\today}

\begin{abstract}
Starting from {\it ab-initio} band structure for Na$_x$CoO$_2$, we
derive the single-electron energies and the effective tight-binding
description for the $t_{2g}$ bands using a projection procedure. We
find that due to the presence of the next-nearest-neighbor hoppings
a local minimum in the electronic dispersion close to the $\Gamma$
point of the first Brillouin zone forms. Therefore, in addition to a
large Fermi surface an electron pocket close to the $\Gamma$ point
emerges at high doping concentrations. The latter yields the new
scattering channel resulting in a peak structure of the itinerant
magnetic susceptibility at small momenta. This indicates itinerant
in-plane ferromagnetic state above certain critical concentration
$x_m$, in agreement with neutron scattering data. Below $x_m$ the
magnetic susceptibility shows a tendency towards the 
antiferromagnetic fluctuations. We estimate the value of 
$0.56 < x_m < 0.68$ within the rigid band model and within the Hubbard
model with infinite on-site Coulomb repulsion consistent with the
experimental phase diagram.
\end{abstract}


\maketitle

\textbf{1.}
The diverse physical properties of the cobaltate Na$_x$CoO$_2$
attracted much attention after the discovery of the unconventional
superconductivity in its hydrated counterpart, Na$_x$CoO$_2 \cdot
y$H$_2$O \cite{kt2003}. The phase diagram of this compound, with
varying electron doping $x$ and water intercalation $y$, is rich and
complicated; in addition to superconductivity, it exhibits magnetic
and charge orders, and some other structural transitions
\cite{it1997,yw2000,mlf2004,bcs2004}. Parent compound,
Na$_x$CoO$_2$,  is a quasi-two-dimensional system with Co in CoO$_2$
layers forming a triangular lattice where Co-Co in-plane distance
three times smaller than the inter-plane one. Na ion resides between
the CoO$_2$ layers and gives additional $x$ electrons to the layer,
lowering Co valence from Co$^{4+}$ ($3d^5$ configuration) to
Co$^{3+}$ ($3d^6$ configuration) upon $x$ changing from 0 for virtual 
compound CoO$_2$ to 1 for NaCoO$_2$.
The hole in the $d$-orbital occupies one of the $t_{2g} $ levels, which
are lower than $e_g$ levels by about 2 eV \cite{djs2000}. The
degeneracy of the $t_{2g}$ levels is partially lifted by the
trigonal distortion which splits it into the higher $a_{1g}$ singlet
and the lower two $e'_g$ states.

First principle LDA (local density approximation) and LDA+U band
structure calculations predict Na$_x$CoO$_2$ has a large Fermi
surface (FS) having mainly $a_{1g} $ character and centered around the
$\Gamma=(0,0,0)$ point and also six hole pockets of mostly $e'_g$
character near the ${\rm K}=(0,\frac{4\pi}{3},0)$ points of the
hexagonal Brillouin zone for a wide range of $x$ \cite{djs2000,
kwl2004}. At the same time, recent Angle-Resolved Photo-Emission
Spectroscopy (ARPES) experiments \cite{mzh2004, hby2004, hby2005,
dq2006} reveal doping dependent Fermi surface evolution for a wide
range of Na concentrations ($0.3 \le x \le 0.8)$ with no sign of the
hole pockets. The observed Fermi surface is centered around the
$\Gamma $ point and have mostly $a_{1g}$ character. Furthermore,
measured dispersion of the top of the valence band is twice as
narrower as compared to the LDA calculated bands.

Concerning magnetic properties the local spin density approach
(LSDA) predicts Na$_x$CoO$_2$ to have a weak intra-plane itinerant
ferromagnetic (FM) state for almost all Na concentrations, $0.3 \leq
x \leq 0.7$ \cite{djs2003}. On the contrary, neutron scattering
finds the A-type antiferromagnetic order implying the ferromagnetic
order within Co-layer {\it only} for $0.75 \leq x \leq 0.9$  with
ordering temperature $T_m \approx 22$ K, and with inter-plane $J_c$
and intra-plane $J_{ab}$ exchange constants to be 12 meV and -6 meV,
respectively \cite{atb2004,spb2005,lmh2005}.

In this letter we derive an effective low-energy model describing
the bands crossing the Fermi level on the basis of the LDA band
structure calculations. Due to the FS topology,
inferred from LDA band structure, the magnetic susceptibility
$\chi_0({\bf q},\omega=0)$ reveals two different regimes for
different dopings: for $x < 0.58$ it shows pronounced peaks at
antiferromagnetic (AFM) wave vector
${\bf Q}_{AFM}=\left\{\left(\frac{2\pi}{3},\frac{2\pi}{\sqrt{3}}\right),
\left(\frac{4\pi}{3},0\right) \right\}$ resulting in the tendency
towards the in-plane $120^\circ$ AFM order, while for $x > 0.58$ the
peaks at small momenta near the ${\bf Q}_{FM}=\left(0,0\right)$
form, clearly demonstrating the tendency of the system towards the
itinerant in-plane FM ordered state. We find that the formation of
the electron pocket around the $\Gamma$ point is crucial for the
in-plane FM ordering at high doping concentrations.

\begin{figure}
\includegraphics[width=0.85\linewidth]{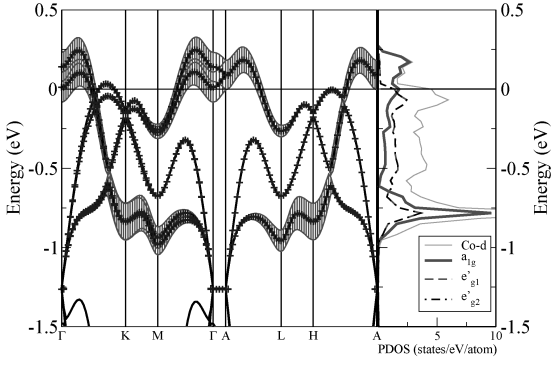}
\caption{Calculated near-Fermi level LDA band
structure and partial density of states (PDOS) for
Na$_{0.33}$CoO$_2$. The contribution of Co-$a_{1g}$ states is
denoted by the vertical broadening of the bands with
thickness proportional to the weight of the contribution. The
crosses indicate the dispersion of the bands obtained by projection
on the $t_{2g}$ orbitals.} \label{fig:fig1}
\end {figure}

\textbf{2.}
The band structure of Na$_{0.33}$CoO$_2$ (Fig.~\ref{fig:fig1}) was
obtained within the LDA \cite{Kohn65} in the framework of
TB-LMTO-ASA (Tight Binding approach to the LMTO using Atomic Sphere
Approximation) \cite{Andersen84} computation scheme. 
The crystal structure parameters were taken from \cite{jorgensen2003}.
The bands crossing the Fermi level have mostly $a_{1g}$ character, consistent
with previous LDA findings \cite{djs2000}. Note, the small FS
pockets near the K point with $e'_g$-symmetry present at $x=0.33$
disappear for higher dopings because the corresponding bands sink
below the Fermi level.

To construct the effective Hamiltonian and to derive the effective
Co-Co hopping integrals $t_{fg}^{\alpha \beta}$ for the
$t_{2g}$-manifold we apply the projection procedure \cite{Marzari97,
Anisimov05}. Here, $(\alpha \beta)$ denotes a pair of orbitals,
$a_{1g}$, $e'_{g1}$ or $e'_{g2}$. The indices $f$ and $g$ correspond
to the Co-sites on the triangular lattice. 
The obtained hoppings are given in the Table~\ref{table:params}, and
the obtained single-electron energies $\varepsilon^\alpha$ are equal to 
(in eV, relative to $\varepsilon^{a_{1g}}$):
$\varepsilon^{a_{1g}}=0$, $\varepsilon^{e'_{g1}}=\varepsilon^{e'_{g2}}=-0.053$.

\begin{figure}
\includegraphics[width=0.85\linewidth]{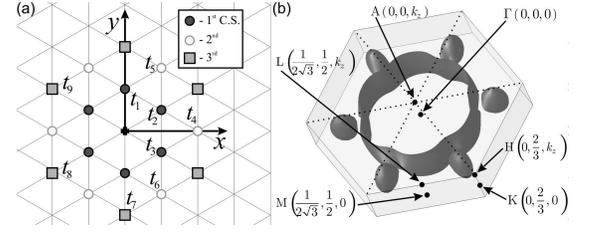}
\caption{(a) Schematic crystal structure of the
Co-layer in Na$_x$CoO$_2$ with hopping notations within the first
three coordination spheres (C.S.). (b) LDA-calculated Fermi surface
with cylindrical part having mostly $a_{1g}$ character and
six hole pockets having mostly $e'_{g}$ character. $k_x$ and
$k_y$ coordinates of the symmetry points are given in units of 
$2\pi/a$ with $a$ being the in-plane lattice constant.} \label{fig:fig2}
\end {figure}

An agreement between the bands obtained using projection procedure
and the LDA bands shown in Fig.~\ref{fig:fig1}, that confirms the
Co-$t_{2g}$ nature of the near-Fermi level bands \cite{djs2000,johannes2004}. 
For simplicity we have enumerated site pairs,
$t_{fg}^{\alpha \beta} \rightarrow t_n^{\alpha \beta}$, with $n=0,
1, 2, ...$ (see Fig.~\ref{fig:fig2}a and the correspondence
between in-plane vectors and index $n$ in the Table~\ref{table:params}).
Due to $C_3$ symmetry of the cobaltate lattice, the following
equalities are present: $| t_3^{\alpha \beta} | = | t_1^{\alpha
\beta} |$, $| t_5^{\alpha \beta} | = | t_4^{\alpha \beta} |$, $|
t_9^{\alpha \beta} | = | t_7^{\alpha \beta} |$. In addition
$t_1^{\alpha \beta} = t_2^{\alpha \beta}$ for $a_{1g} \rightarrow
a_{1g}$ hoppings, which, however, does not hold for $e'_{g1,2}$
orbitals. Thus, since the hybridization between the $a_{1g}$ and the
$e'_{g}$ bands is not small, a simplified description of the bands
crossing the Fermi level in terms of the $a_{1g}$ band only
(neglecting $e'_{g}$ band and corresponding hybridizations, see e.g.
\cite{kk2005}) may lead to a incorrect result due to its higher
symmetry. In the following we neglect the inter-layer splitting
present for $k_z=0$ plane because of its subtle effect on the
topology of the FS \cite{djs2000}.
\begin{table*}
\caption{The in-plane hopping integrals $t_n^{\alpha \beta}$ for different in-plane vectors $n=(f,g)$ for Na$_x$CoO$_2$, 
$x = 0.33$.}
\label{table:params}
\begin{ruledtabular}
\begin{tabular}{c|c|c|c|c|c|c|c|c|c}
$n=(f,g)$: & (0, 1) & ($\frac{\sqrt{3}}{2}$, $\frac{1}{2}$) & ($\frac{\sqrt{3}}{2}$,
-$\frac{1}{2}$) & ($\sqrt{3}$, 0) & ($\frac{\sqrt{3}}{2}$, $\frac{3}{2}$) 
& ($\frac{\sqrt{3}}{2}$, -$\frac{3}{2}$) & (0, 2) & ($\sqrt{3}$, 1) & ($\sqrt{3}$, -1) \\
\hline
$\alpha \rightarrow \beta$
& $t_1^{\alpha \beta}$ & $t_2^{\alpha \beta}$ & $t_3^{\alpha \beta}$ & $t_4^{\alpha \beta}$
& $t_5^{\alpha \beta}$ & $t_6^{\alpha \beta}$ & $t_7^{\alpha \beta}$ & $t_8^{\alpha \beta}$ & $t_9^{\alpha \beta}$ \\
\hline \hline
$a_{1g}  \rightarrow a_{1g}$  &  0.123& 0.123&  0.123&-0.022& -0.022&-0.021&-0.025 &-0.025& -0.025 \\
$a_{1g}  \rightarrow e'_{g1}$ & -0.044& 0.089& -0.044& 0.010&  0.010&-0.021& -0.021& 0.042& -0.021 \\
$a_{1g}  \rightarrow e'_{g2}$ & -0.077& 0.000&  0.077& 0.018& -0.018& 0.000& -0.036& 0.000&  0.036 \\
$e'_{g1} \rightarrow e'_{g1}$ & -0.069&-0.005& -0.069& 0.018&  0.018&-0.026&-0.017 &-0.085& -0.017 \\
$e'_{g1} \rightarrow e'_{g2}$ &  0.037& 0.000& -0.037&-0.026&  0.026& 0.000&-0.039 & 0.000&  0.039 \\
$e'_{g2} \rightarrow e'_{g2}$ & -0.026&-0.090& -0.027&-0.011& -0.011& 0.033&-0.062& 0.006 & -0.062
\end{tabular}
\end{ruledtabular}
\end{table*}

Then the free electron Hamiltonian for CoO$_2$-plane in a hole
representation is given by:
\begin{equation}
H_0 = - \sum\limits_{{\bf k},\alpha ,\sigma } {\left( {\varepsilon
^\alpha - \mu } \right)n_{{\bf k} \alpha \sigma } } -
\sum\limits_{{\bf k}, \sigma} \sum\limits_{\alpha, \beta}
t_{{\bf k}}^{\alpha \beta } d_{{\bf k} \alpha \sigma }^\dag d_{{\bf k} \beta \sigma}.
\label{eq:H0}
\end{equation}
where $d_{{\bf k} \alpha \sigma}$ ($d_{{\bf k} \alpha \sigma}^\dag$) is the annihilation (creation) operator
for hole with momentum ${\bf k}$, spin $\sigma$ and orbital index $\alpha$,
$n_{{\bf k} \alpha \sigma} = d_{{\bf k} \alpha \sigma}^\dag d_{{\bf k} \alpha \sigma}$,
and $t_{{\bf k}}^{\alpha \beta }$ is the Fourier transform of the hopping matrix element.
Introducing matrix notations, $\left( {\hat {t}_{{\bf k}} } \right)_{\alpha \beta} = t_{{\bf k}}^{\alpha \beta}$ and
$\left( {\hat {t}_n } \right)_{\alpha \beta} = t_n^{\alpha \beta}$,
the hoppings matrix elements in the momentum representation are given by:
\begin{eqnarray}
\hat{t}_{{\bf k}} &=& 2 \hat{t}_1 \cos k_2 + 2 \hat{t}_2 \cos k_3 + 2 \hat {t}_3 \cos k_1 + \nonumber \\
&+& 2 \hat{t}_4 \cos (k_1 + k_3) + 2 \hat{t}_5 \cos (k_2 + k_1) + 2 \hat{t}_6 \cos (k_1 - k_2) + \nonumber \\
&+& 2 \hat{t}_7 \cos 2 k_2 + 2 \hat{t}_8 \cos 2 k_3 + 2 \hat{t}_9 \cos 2 k_1,
\end{eqnarray}
where $k_1 = \frac{\sqrt 3}{2} k_x - \frac{1}{2} k_y$,  $k_2 = k_y$,
$k_3 = \frac{\sqrt 3}{2} k_x + \frac{1}{2} k_y$.

\begin{figure}
\includegraphics[width=0.9\linewidth]{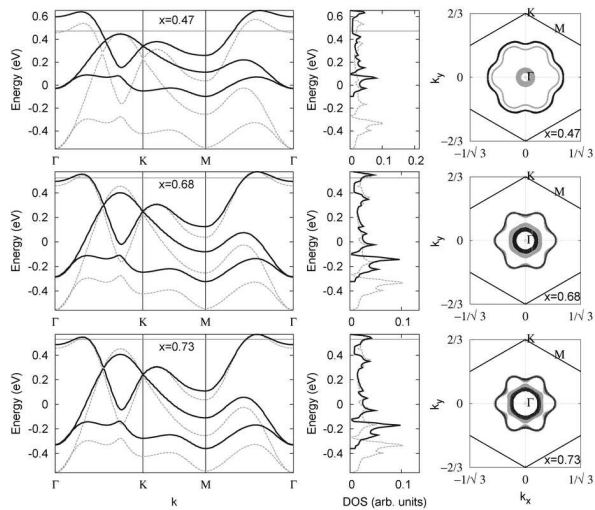}
\caption{Calculated band structure and Fermi surface
topology for Na$_x$CoO$_2$ for $x = 0.47, 0.68, 0.73$. The dashed
(light gray) and solid (black) curves represent the rigid-band approximation
and the Hubbard-I solution, respectively. The horizontal
line denotes self-consistently calculated chemical potential $\mu$.}
\label{fig:fig3}
\end {figure}

Within this rigid band approximation the doping-dependent evolution
of the electronic dispersion, the density of states (DOS) and the FS
is shown in Fig.~\ref{fig:fig3}. We notice, that already at
$x=0.48$ the FS $e'_g$ hole pockets are absent. Most importantly we
find another interesting feature. Namely, the local minimum of the
band dispersion around the $\Gamma$ point yields the appearance of
the second FS contour centered around this point. This electron FS
pocket becomes larger upon increasing doping $x$. As was shown in
Ref.~\cite{brenig2004} for the Hubbard model on a triangular
lattice, the main reason for the local minimum around the $\Gamma$
point is the presence of the next-nearest-neighbor hoppings, which
also enter in our calculations. Although this minimum is not yet
directly observed in ARPES experiments, the presence of the
associated second FS contour would reduce the FS volume and resolve
an issue why the volume of the FS observed in ARPES is larger than
it is expected from Luttinger's theorem \cite{geck2006}.
Furthermore, an emergence of this pocket would influence the Hall
conductivity at high doping concentrations which is interesting to
check experimentally.

\begin{figure}
\includegraphics[width=0.85\linewidth]{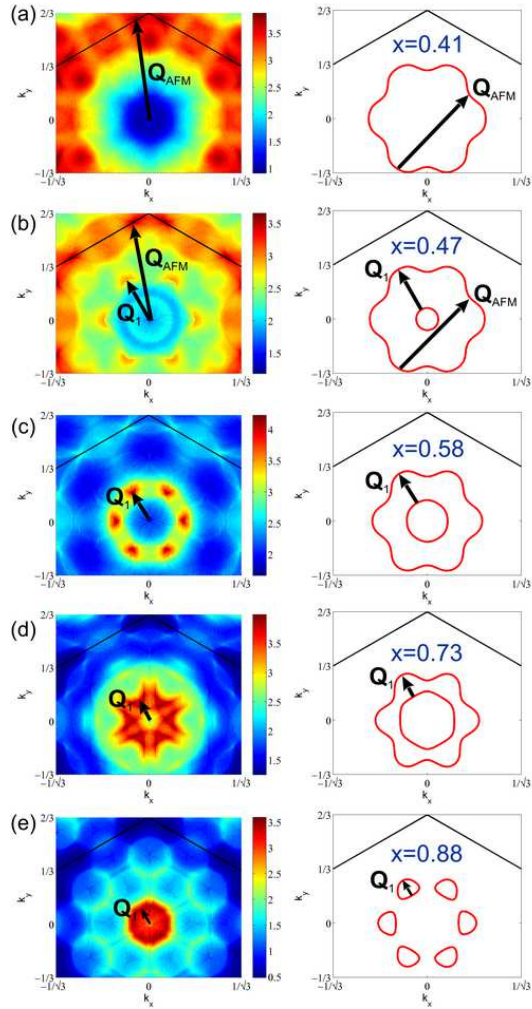}
\caption{The grayscale plot of the real part of the
magnetic susceptibility ${\rm Re} \chi_0 ({\bf k},\omega=0)$ as a
function of the momentum in units of $2\pi / a$ (left), and the
Fermi surface for corresponding doping $x$ (right). 
The arrows indicate the scattering wave vectors ${\bf Q_i}$ 
as described in the text.}
\label{fig:fig4}
\end {figure}

\textbf{3.}
To analyze a possibility of the itinerant magnetism we calculate the
magnetic susceptibility $\chi_0({\bf q},\omega=0)$ based on the
Hamiltonian $H_0$. The doping-dependent evolution of the peaks in
${\rm Re} \chi_0({\bf q},0)$ is shown in Fig.~\ref{fig:fig4}. At
$x=0.45$ the $e'_g$ bands are below the Fermi level, and the FS has
the form of the rounded hexagon. It results in a number of nesting
wave vectors around the antiferromagnetic wave vector
${\bf Q}_{AFM}$. The corresponding broad peaks in the ${\rm Re}
\chi_0({\bf q},0)$ appear around ${\bf Q}_{AFM}$, indicating the
tendency of the electronic system towards the $120^\circ$ AFM
ordered state \cite{mdj2004}. Upon increasing doping, the Fermi level 
crosses the local minimum at the $\Gamma$ point, resulting in the second 
almost circle FS contour. As soon as this change of the FS topology
takes place, the scattering at the momentum ${\bf Q}_{AFM}$ is
quickly suppressed, and vanishes already at $x_m \approx 0.56$. Most
importantly, a new scattering vector, ${\bf Q}_1$, appears. 
This wave vector is small and yields peaks in the magnetic susceptibility 
at small momenta, indicating the tendency of the magnetic system to shift 
towards itinerant FM order.
For larger $x$ the inner FS contour increases leading to a further
decrease of the ${\bf Q}_1$. In the case of $x \approx 0.88$ the
FS topology changes again, resulting in a six distant FS contours,
yielding even smaller length of ${\bf Q}_1$.
The obtained scattering at small momenta  in the magnetic susceptibility
for $x > x_m$ is qualitatively consistent with the scattering around
${\bf Q}_{FM}=\left(0,0\right)$, observed in the neutron scattering experiments
\cite{atb2004,spb2005,lmh2005}.

\textbf{4.}
Since obtained magnetic susceptibility depends mostly on the
topology of the FS one expects that the behavior shown in Fig.~\ref{fig:fig4} 
will be valid even if one includes an interaction
term $H_{int}$ into account, at least in the case if it is the
on-site Hubbard interaction $U$. The only difference would be a
shift of the critical concentrations $x_m$, at which the FS topology
changes and tendency to the AFM order changes towards the FM ordered
state. To check this, we have taken the strong electron correlations
into account by adding the on-site Coulomb interaction terms to the
$H_0$, similar to Refs. \cite{sz2005, mi2005}. The effective on-site
Hubbard repulsion $U_{eff} \approx 4$ eV on Co sites is much larger
than the bare bandwidth $W \approx 1.2$ eV, and, thus, it is
possible to project doubly occupied states out and formulate an
effective model equivalent to the Hubbard model with infinite value
of $U$.

In the atomic limit local low-energy states on the Co sites are the
vacuum state $\left| 0 \right\rangle $ and the single-hole states
$\left| {a \sigma } \right\rangle $, $\left| {e_1 \sigma }
\right\rangle $, $\left| {e_2 \sigma } \right\rangle$. The simplest
way to describe the quasiparticle excitations between these states
is to use the projective Hubbard $X$-operators \cite{hubbard1964}:
$X_f^m \leftrightarrow X_f^{p,q} \equiv \left| p \right> \left< q \right|$, 
where index $m \leftrightarrow (p,q)$ enumerates
quasiparticles. There is a simple correspondence between
$X$-operators and single-electron creation-annihilation operators:
$d_{f\alpha \sigma } = \sum\limits_m { \gamma_{\alpha \sigma}(m) X_f^m }$, 
where $\gamma_{\alpha \sigma }(m)$ determines the partial
weight of a quasiparticle $m$ with spin $\sigma$ and orbital index
$\alpha$. In these notations the Hamiltonian of the Hubbard model in
the limit $U \to \infty$ has the form:
\begin{eqnarray}
H = - \sum\limits_{f,p} {\left( {\varepsilon^p - \mu } \right)X_f^{p,p} }
- \sum\limits_{f \ne g} {\sum\limits_{m,m'} {t_{fg}^{mm'} {X_f^{m}}^\dag X_g^{m'} } }.
\end{eqnarray}
In the so-called Hubbard-I approximation within the generalized
Dyson equation for the $X$-operators \cite{zaitsev1975,izumov1991,ovchinnikov_book2004} 
the quasiparticle bands formed by the $a_{1g} \to a_{1g}$ hoppings will
be renormalized by the $(1 + x)/2$ factor, while the quasiparticle
bands formed by the $e'_g$ hoppings will be renormalized by $x$.

In Fig.~\ref{fig:fig3} the quasiparticle spectrum is shown. 
One finds within Hubbard-I approximations the bands
become narrower with lowering $x$ due to doping dependence of the
quasiparticle's spectral weight. Most importantly, the doping
evolution of the FS is qualitatively the same as in the rigid-band
picture. Thus, a bandwidth reduction and a spectral weight
renormalization do not change the topology of the FS. Therefore, the
presence of the strong electronic correlations do not change
qualitatively our results for the bare susceptibility.
Quantitatively, the critical concentration $x_m$ shifts towards
higher values and within Hubbard-I it becomes $x_m \approx 0.68$.

\textbf{5.}
To conclude, we have shown that in the model with {\it ab-initio}
calculated parameters the magnetic susceptibility is doping
dependent. At the critical doping concentration, $x_m$, the electron
pocket on the FS in the center of the Brillouin zone well develops.
For $x < x_m$, the system shows tendency towards the $120^\circ$ AFM
ordered state, while for $x > x_m$ the peak in the magnetic
susceptibility is at small wave vectors indicating strong tendency
towards the itinerant FS state. Within the tight-binding model $x_m$
is estimated to be around $0.56$. Analyzing the influence of the
strong Coulomb repulsion and the corresponding reduction of the
bandwidth and the quasiparticle spectral weight in the
strong-coupling Hubbard-I approximation, we show that the critical
concentration changes to the $x_m \approx 0.68$. At the same time,
the underlying physics of the formation of the itinerant FM state
remains the same.

We would like to thank G. Bouzerar, W. Brenig, P. Fulde, 
S.G. Ovchinnikov, D. Singh, and Ziqiang Wang for useful discussions, 
I. Mazin and N.B. Perkins for the careful reading of the manuscript, 
and S. Borisenko for sharing with us the experimental
results prior to publication. The work of M.M.K. was supported by
INTAS YS (Grant 05-109-4891) and RFBR (Grants 06-02-16100 and 06-02-90537-BNTS).
A.S. and V.I.A. acknowledge the financial support from RFBR (Grants
04-02-16096, 06-02-81017), and NWO (Grant 047.016.005).

\end{document}